# A scalable noisy speech dataset and online subjective test framework

*Chandan K. A. Reddy[1], Ebrahim Beyrami[1], Jamie Pool[1], Ross Cutler[1], Sriram Srinivasan[1], Johannes Gehrke[1]*

[1]Microsoft Corp., USA

{Chandan.Karadagur, ebbeyram, Jamie.Pool, Ross.Cutler}@microsoft.com,
{Sriram.Srinivasan, Johannes}@microsoft.com

## Abstract

Background noise is a major source of quality impairments in Voice over Internet Protocol (VoIP) and Public Switched Telephone Network (PSTN) calls. Recent work shows the efficacy of deep learning for noise suppression, but the datasets have been relatively small compared to those used in other domains (e.g., ImageNet) and the associated evaluations have been more focused. In order to better facilitate deep learning research in Speech Enhancement, we present a noisy speech dataset (MS-SNSD) that can scale to arbitrary sizes depending on the number of speakers, noise types, and Speech to Noise Ratio (SNR) levels desired. We show that increasing dataset sizes increases noise suppression performance as expected. In addition, we provide an open-source evaluation methodology to evaluate the results subjectively at scale using crowdsourcing, with a reference algorithm to normalize the results. To demonstrate the dataset and evaluation framework we apply it to several noise suppressors and compare the subjective Mean Opinion Score (MOS) with objective quality measures such as SNR, PESQ, POLQA, and VISQOL and show why MOS is still required. Our subjective MOS evaluation is the first large scale evaluation of Speech Enhancement algorithms that we are aware of.

## 1. Introduction

Background noise is a major source of quality impairments in Voice over Internet Protocol (VoIP) and Public Switched Telephone Network (PSTN) calls. In a major consumer VoIP provider (Skype) background noise is a top-5 cause for calls to be rated poor. In conventional Speech Enhancement (SE) techniques, the parameters required for noise suppression are derived based on statistical models and are estimated from the noisy observations. These algorithms generalize well to unseen acoustic conditions but fail to significantly reduce non-stationary noise types as we show in this paper. This has motivated researchers to explore deep learning techniques for noise suppression. While there have been recently published noise suppression algorithms using Deep Neural Networks (DNN)(e.g., [1]–[3]), the datasets used are relatively small compared (e.g., 23K clips in [4]) to those used in other domains (e.g., ImageNet [5] with 14M annotated images). A result of using relatively small datasets is insufficient coverage of various noisy speech conditions, which will result in poor performance on the test data with unseen noisy speech conditions. There has been less work done in this area of DNN based SE algorithms trained on a limited training set [6] and methods to scale up the dataset [7]. The literature shows that having more training data improves performance [6] using objective measures. Another shortcoming of having a fixed number of noisy-clean speech pairs is that different DNN based techniques require different amounts of data to achieve a certain performance threshold. Hence, there is a need for scalable data sets with good coverage of noisy conditions such as different speakers, noise types, and Speech to Noise Ratios (SNRs).

Most of the published SE methods are evaluated using objective measures such as Perceptual Evaluation of Speech Quality (PESQ) [8], Perceptual Objective Listening Quality Analysis (POLQA) [9] and Virtual Speech Quality Objective Listener (ViSQOL) [10]. Subjective tests are typically conducted by measuring Mean Opinion Score (MOS) [11]. The objective metrics are supposed to predict MOS with high correlation. However, the reliability of these objective metrics is under question as we show subsequently that they correlate poorly with MOS. Hence, MOS is still the most reliable metric for speech quality. The downside of conducting offline subjective experiments is that it is not scalable to evaluate large datasets as it is expensive and time-consuming. This is the primary reason why the papers that report MOS evaluate on a very small evaluation dataset.

In this paper, we introduce the Microsoft Scalable Noisy Speech Dataset (MS-SNSD) and evaluate three state of the art and popular DNN based SE techniques; we also include the Wiener filter as the baseline for comparison. The three DNN based methods considered are i) SEGAN [3]; ii) A Wavenet for Speech Denoising [2]; and iii) an improved version of RNNoise [1]. All these methods were retrained on MS-SNSD to give a fair comparison. The performance of these methods was evaluated using the proposed online subjective evaluation framework to measure MOS. We also evaluate the performance using the objective metrics and measure their correlation with MOS.

The proposed subjective evaluation framework uses online tools such as Microsoft's Universal Human Relevance System (UHRS) or Amazon's Mechanical Turk [12], where the tasks are crowdsourced to 1000s of click/hit workers online. These tools are being used extensively to get annotations and ratings for large datasets in many fields to train DNN models. We lay out a procedure to develop a reliable Hit Application for conducting a subjective evaluation. The proposed framework is used to evaluate the three DNN based SE methods with Wiener filter and the original noisy speech on large-scale test data that covers a variety of noisy speech conditions. A total of 27,500 audio clips were used for subjective tests, which to our knowledge, is the largest subjective evaluation conducted for a speech enhancement task.

## 1.1. Related work

The Noizeus dataset [13] is a widely used narrowband dataset with about 0.7 hours of clean and noisy speech audio clips. The Edinburgh dataset [4] is a wideband dataset with 16 hours of clean and noisy audio clips. The Edinburgh dataset doesn't use data augmentation for noise clips and SNR levels, so the number of audio clips are:

$$N_1 = speakers \cdot sentences$$

Data augmentation has been shown to be very effective in deep learning models, and our dataset uses data augmentation on noise clips and SNR levels:

$$N_2 = speakers \cdot sentences \cdot noise \cdot SNR$$

For example, with 14 noise sources and 5 SNR levels the dataset size is increased by 70 times compared to without using data augmentation on noise and SNR levels.

## 1.2. Contributions

This paper makes the following contributions:

- We provide an open-source scalable noisy speech dataset that provides orders of magnitude more training data by augmentation. All source is available to extend the dataset[1]. All sound files have favorable reuse licenses.
- We provide an open-source framework for online subjective evaluations of the results with references that allow future comparisons.
- We show the best performing noise suppressor, RNNoise, significantly improves with larger datasets.
- We provide a comparison of MOS with 4 noise suppressors, which is the most extensive subjective comparison of noise suppressors we are aware of.
- We give a breakdown of performance by noise types which show which areas need the most work.
- We provide the correlation of MOS and SNR, PESQ, POLQA, VISQOL which shows why MOS is required for noise suppression comparisons.

## 2. Dataset

The noisy speech dataset is generated to satisfy the following needs:

- **Scalability** – The dataset should be scalable as a function of the number of speakers, noise types and SNR levels desired. The dataset should be able to easily accommodate new noisy conditions as required.
- **Flexibility** – Having flexibility in setting the desired lengths of each audio clip used for training and setting the sampling rate. Having flexibility in segregating the audio clips based on their noise types and SNR levels. This is important as different DNN based SE techniques have different requirements. For example, RNNoise learns from the sequence information and hence it is of interest to have input clips with similar noise types and SNR levels.

[1] Source code and data is available at http://aka.ms/ms-snsd

## 2.1. Clean Speech Corpus

### 2.1.1. Clean Speech for Training

We used the Edinburgh 56 speaker clean speech dataset (28 male and 28 female speakers) [4] for training the models. This dataset has over 23K clips of speakers reading a short sentence (median length 2.6 seconds), all resampled to 16 kHz and single channel.

### 2.1.2. Clean Speech for Testing

For testing and subjective evaluations, we used the Graz University's clean speech dataset which consists of 4,720 recorded sentences spoken by 20 speakers. The clean speech clips for testing were concatenated to form a clip length of 10 secs on an average.

## 2.2. Noise database

The noise clips were selected from DEMAND database [14] and Freesound.org. We hand-picked the clips by carefully listening to ensure the quality of the recordings. The chosen noise types were relevant to our application, but it can always be scaled to accommodate new types. The 14 noise types chosen are: air conditioner, announcements, appliances (washer/dryer), car noise, copy machine, door shutting, eating (munching), multi-talker babble, neighbor speaking, squeaky chair, traffic, road, typing, vacuum cleaner. 10 unique and diverse recordings were selected for each of these noise types.

For the test dataset, the noise clips chosen were different recordings than that of the training set, but from similar noise categories as that of the training set.

## 2.3. Microsoft Scalable Noisy Speech Database (MS-SNSD)

The framework for noisy speech generation takes in the directories pointing to the clean speech and noise as inputs. Other parameters that can be controlled are the desired number of noisy speech clips, length of the clip and SNR levels. Based on these parameters, the clean utterances from a speaker are randomly chosen and concatenated to form the desired length. Similarly, the noise segment with the same length of clean speech is randomly chosen and added to the noise at different SNR levels. All the files are sampled at 16 kHz and normalized to -25 dBFS. In this work, we used a total of 50,000 noisy speech clips (42 hours) to train the DNN models.

The total number of test clips were 5500 (16 hours), which is the most extensive test set for subjective tests that we are aware of.

## 3. Online subjective evaluation framework

### 3.1. Hit Application for online subjective evaluation

In this work, we used Microsoft's UHRS tool for conducting online subjective experiments. The Hit Application is the framework written in Javascript to create the User Interface (UI) and control other parameters. The designed UI is shown in Figure 1. The click workers/judges are compensated based on the number of jobs they complete. The judges are expected to carefully listen to each audio clip and rate the clip based on the quality of the speech perceived. The subjective studies can also be done with Mechanical Turk (source code is provided).

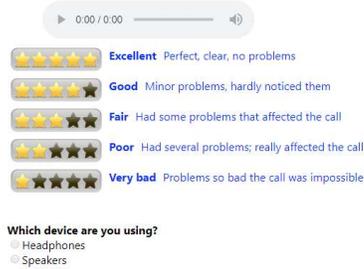

Figure 1: *UHRS Hit Application*

### 3.2. Quality control

Every judge must read the guidelines, go through the training and pass the qualification test before they are permitted to start the main experiments. In the training phase, the click workers are provided with 5 clips that cover both clean speech and noisy speech. They listen to these clips and get familiar with the expectations and motive of the experiment. During the qualification test, the judges are provided with clips that are either clean speech (5 stars) or extremely noisy speech (1 star). The expected rating is either 1 or 5, and judges are expected to get at least 80% right to pass the qualification test. This is to ensure that the click workers have a playback device (speaker or headphone) and they are not hearing impaired. Spam check is enabled to monitor click workers who are consistently off from the mean by a significant margin. These click workers are prevented from taking further tests.

### 3.3. Using reference SE methods

The noisy clips and a reference SE method, Wiener Filter, are used to normalize future SE evaluations. Specifically, future SE MOS results are linearly scaled to match the reference Noisy and Wiener values provided in Table 1. A utility script is provided in the source code to do this normalization. The full source of the Wiener Filter is also provided.

## 4. Overview of the SE methods compared

A brief introduction of the methods used for comparison is given below. All these methods are already published and gaining traction in the Speech Enhancement community.

### 4.1. Wiener Filter

Wiener filter works on Minimizing the Mean Squared Error (MMSE) between the estimated clean speech magnitude spectrum $\hat{S}(k)$ and the original clean speech magnitude spectrum $S(k)$, where $k$ is the frequency bin. The optimum filter is given by [15][16],

$$H(k) = \frac{P_s(k)}{P_s(k)+P_n(k)} \quad (1)$$

where $P_s(k)$ and $P_n(k)$ represent the estimated power spectral densities of the clean speech and noise respectively. The clean speech estimate is given by,

$$\hat{S}(k) = X(k)H(k) \quad (2)$$

where $X(k)$ is the original noisy speech spectrum. The time series is reconstructed using the noisy phase.

The magnitude spectrum is computed on frames of length 20 ms with 50% overlap. The noise power spectra $P_n(k)$ is computed during noise-only frames using an energy threshold-based Voice Activity Detector (VAD).

### 4.2. Improved RNNoise

RNNoise [1] is a low-complexity noise suppression algorithm using recurrent neural networks. Instead of estimating raw PCM samples, the method estimates noise suppression gains in relevant critical bands. The original method operates at a sampling rate of 48 kHz, which we modified to operate for wideband speech at 16 kHz and adjusted the critical bands accordingly. We also limited the amount of attenuation to approximately 15 dB, which we observed resulted in a subjectively better trade-off between speech distortion and noise reduction. RNNoise, with these modifications, was retrained on the 42-hour dataset described in section 2.3.

### 4.3. WaveNet

WaveNet [2] provides an end-to-end learning model for speech denoising. The proposed structure takes a series of non-casual convolutional filters with exponentially increasing dilation factors and predicts target fields. The supervised model learns by minimizing an energy-conserving loss function.

### 4.3. SEGAN

SEGAN [3] is another end-to-end learning model that works with raw audio data. The training algorithm defines two networks a generative network that maps the noisy speech into clean speech and a discriminative network which distinguishes whether inputs come from clean or enhanced speech. The learning process is defined as a minmax game between the two networks. The generative network for SEGAN is a fully convolutional network and the discriminator can be viewed as a trainable loss function.

## 5. Experimental results

### 5.1. Performance analysis

The methods described in Sections 4.2, 4.3 and 4.4 were trained using the generated MS-SNSD dataset. The test dataset consisted of 5,500 noisy speech clips that covered various noisy conditions, which were enhanced by 4 methods. In total, there were 27,500 audio clips rated by 10 judges per clip. MOS is the average of the ratings by 10 judges. The average of the MOS results with confidence intervals for each of these methods is shown in Table 1. Figure 2 shows the comparison of MOS distribution. The distribution drifts towards the right if the SE is better.

Among the methods compared, RNNoise clearly outperforms all the other noise suppressors. Though SEGAN is good at suppressing the background noise, it also tends to distort the speech resulting in lower MOS. WaveNet does not show statistically significant improvements either.

Figure 3 shows the impact of each noise type on the MOS for different noise suppressors. The characteristics of all these noise types are different. For example, air conditioner, vacuum cleaner, and copy machine are stationary in nature, but other noise types such as announcements and shutting door are non-stationary. The performance of the compared noise suppressors varies with the noise types. Wiener filter shows decent improvements when the noise is stationary. RNNoise performs well across all noise types. WaveNet performs well on typing noise, while it underperforms on all other types. The graph for the original noisy speech tells us which noise type is more annoying to listeners. Babble, neighbor speaking, and announcements have the lowest MOS, which shows that these noise types should be addressed with special attention. Moreover, the users are more commonly exposed to these three noises types and hence impacts the overall quality ratings in real-world applications.

**5.2. Comparison of MOS with other objective measures**

Many researchers in the area of SE rely on objective metrics like SNR, PESQ, and POLQA to evaluate their methods. However, we show that the correlation of these objective metrics with subjective MOS is not high enough to be reliable for many applications. Table 2 shows the correlation of MOS with PESQ, POLQA, ViSQOL, and SNR. POLQA gives the highest correlation of 0.78, which is still too low to accurately track the quality of the enhanced speech.

Table 1: *Comparison of noise suppression methods on the proposed dataset*

| Noise suppressor | MOS |
|---|---|
| Noisy | 2.45 ± 0.02 |
| SEGAN | 2.17 ± 0.02 |
| WaveNet | 2.39 ± 0.01 |
| Wiener | 2.45 ± 0.02 |
| RNNoise | 2.97 ± 0.02 |

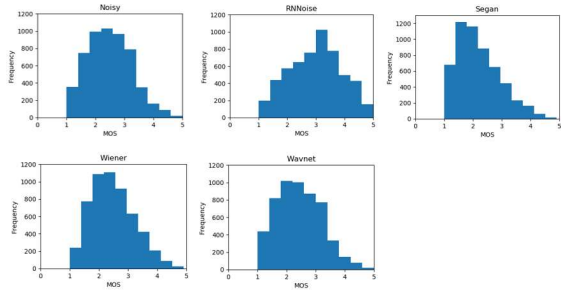

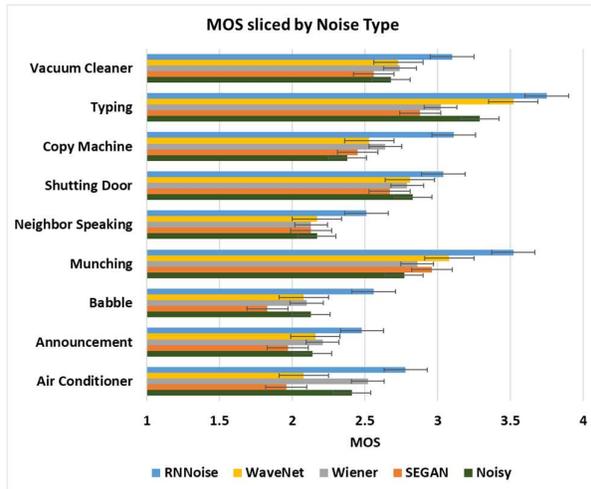

Figure 2: *Distribution of MOS results*

Figure 3: *MOS sliced by Noise Type*

**5.3. Impact of dataset size on performance**

To analyze the impact of the size of the training data set on performance, we trained RNNoise using a 16-hour set (comparable to the size of the Edinburgh set) and a 42-hour training set. The two models were evaluated using an independent test data set that included speech degraded by babble, typing, street and white noise at SNRs of 5 and 15 dB. Table 3 shows the loss corresponding to each data set. For denoising, the loss function is the mean squared error between the ideal and predicted per-band gain function raised to an exponent as defined in [1] eq 8. For voice activity detection (VAD), the loss function in cross-entropy. It is evident that there is a significant reduction in both the denoising and VAD loss when training on the larger data set.

Table 2: *Correlation of MOS with objective measures*

| Objective measures | Pearson Correlation |
|---|---|
| PESQ | 0.75 |
| POLQA | 0.78 |
| VISQOL | 0.61 |
| SNR | 0.56 |

Table 3: *Improvement of RNNoise loss due to an increase in training size using data augmentation*

| Training set hours | Denoising loss | VAD loss |
|---|---|---|
| 16 | 2.87 | 0.24 |
| 42 | 1.78 | 0.16 |

## 6. Conclusions

In this paper, we have provided a scalable noisy speech dataset that allows researchers to create datasets of arbitrary size based on data augmentation and shown the utility of using this dataset to evaluate several noise suppression algorithms. We have also provided an open-source framework that allows researchers to evaluate background noise suppression methods subjectively using crowdsourcing with references, allowing accurate comparisons of SE algorithms between researchers.

For the next steps we are extending the augmentation to include:

- Variances in audio capture sample rates
- Room response transfer functions (reverb)
- Device capture transfer functions

These augmentations will make the dataset more realistic. In addition, we are capturing a large real (not synthetic) test dataset that includes these types of impairments. We believe these additional data augmentations and test datasets will further improve the performance of deep learning-based noise suppression methods, and ultimately improve the end-to-end subjective quality of VoIP and PSTN calls when utilized.

In addition, we will compare using P.835 [17] with P.800 which may lower the MOS variance but at the cost of additional time per ratings.